\documentclass[twocolumn]{article}
\usepackage[left=1.5cm, right=1.5cm, top=1.785cm, bottom=2.0cm]{geometry}
\usepackage{graphicx, hyperref, authblk, bm, amsmath, balance}

\renewcommand{\d}{\text{d}}

\title{Legendre-Fenchel transforms capture layering transitions in porous media}

\author[1]{Olav Galteland}
\author[2]{Eivind Bering}
\author[1]{Kim Kristiansen}
\author[1]{Dick Bedeaux}
\author[1]{Signe Kjelstrup}
\affil[1]{PoreLab, Department of Chemistry, Norwegian University of Science and Technology}
\affil[2]{PoreLab, Department of Physics, Norwegian University of Science and Technology}

\begin{document}
\maketitle
\begin{abstract}
We have investigated the state of a nanoconfined fluid in a slit pore in the canonical and isobaric ensembles. The systems were simulated with molecular dynamics simulations. The fluid has a transition to a close-packed structure when the height of the slit approaches the particle diameter. The Helmholtz energy is a non-convex function of the slit height if the number of particles does not exceed that of one monolayer. As a consequence, the Legendre transform cannot be applied to obtain the Gibbs energy. The Gibbs energy of a non-deformable slit pore can be transformed into the Helmholtz energy of a deformable slit pore using the Legendre-Fenchel transform. The Legendre-Fenchel transform corresponds to the Maxwell construction of equal areas.
\end{abstract}

\section{Introduction}

Over the last years, there has been an increasing number of observations of phase transitions in confined fluids. Fluids can for instance change their critical temperature by several tens of degrees \cite{brennan2002phase, brennan2003molecular}, a two-dimensional layer at an interface may develop more than one structure \cite{galteland2021nanothermodynamic}, and adsorption to droplets may depend largely on droplet size \cite{strom2020thermodynamic, strom2021adsorption}. Classical Gibbs thermodynamics ceases to exist on the nanoscale. The need for the inclusion of shape and size has been met in several ways. Gibbs and followers included curvature as a variable \cite{wilhelmsen2015tolman} to deal with droplet size dependence. Not only size and shape will matter for the outcome of the analysis of simulations; the thermodynamic properties will also depend on the small system's environment or the set of variables that control the system (the ensemble) according to Hill \cite{Hill1963, Hill1964}. Dong \cite{dong2021thermodynamics} argued that thermodynamic variables, like the surface tension, change as we shrink the small system, and proposed to add as variable the integral surface tension, to complement the normal (differential) surface tension. The pressure of fluids in porous media is of special interest as its gradient is the main driving force for mass transport \cite{kjelstrup2018non, kjelstrup2019non}.

A systematic way to address these problems was given by Hill \cite{Hill1963, Hill1964} already 50 years ago. We have argued that the problems are best addressed by his method \cite{Bedeaux2020, galteland2019pressures, erdHos2020gibbs, galteland2021nanothermodynamic}, because the method provides a general description of small systems. We have for instance been able to write scaling laws for small system variables \cite{rauter2020two}, and a new equilibrium criterion for pressure was developed for two-phase equilibria in slit pores \cite{rauter2020two}. Hill's method is therefore our first choice when the aim is to learn more about structural transitions in confined fluids or how variables change. We shall find here that a transforming procedure exists in terms of the Legendre-Fenchel transform. This is a more general transform than the Legendre transform and can be used for large as well as small systems. It will enable us to compute the Gibbs energy from the Helmholtz energy and vice versa. 

We have recently reported the changes in free energy during polymer stretching. The free energy depends on the conditions used, whether the polymer is stretched at controlled length or force \cite{bering2020entropy, bering2020legendre}. For sufficiently short polymers, the Helmholtz energy is a non-convex function of the controlled length of the polymer. A convex function has a non-negative second derivative everywhere. To transform from the Helmholtz to the Gibbs energy the Legendre-Fenchel transform could be applied. Similarly, we observed that the grand potential of a fluid in a slit pore is a non-convex function of the distance between the parallel plates at constant chemical potential and temperature \cite{galteland2021nanothermodynamic}. Also, the Helmholtz energy of solid colloids in solution is non-convex as a function of the controlled distance between the colloids, keeping the temperature, volume, and the number of particles constant \cite{galteland2020solvent}. Both cases can be explained by a disjoining pressure (also known as the solvation pressure) \cite{israelachvili2015intermolecular}, which is the excess normal pressure relative to the bulk pressure due to the packing of fluid particles between the solids. The observations mentioned all stem from size effects. To obtain the free energy of the corresponding constant pressure ensemble the Legendre-Fenchel transform must be applied, and not the Legendre transform.  

The three pillars that science progresses from are theory, experiments, and simulations. The theory part is lacking in nanotechnology. Energy converting devices are abundant, but there is little available general theory of energy conversion for the nanoscale. The laws of energy conversion are the laws of thermodynamics, and the question we are asking is which form these take. Nanothermodynamics has been constructed, mostly by adding terms to Gibbs's classical formulation for large-scale systems. While this mending procedure may serve the purpose in some special cases, it does not present us with a systematic procedure to be used as a general tool. Here we argue that the theory of Hill presented more than 50 years ago, presents an underused opportunity for a systematic procedure. To show the advantage of this approach, we study a transition between two structural regimes in a molecular fluid in a porous media model and document the applicability of an important tool, namely the Legendre-Fenchel transform.

Several problems arise for systems dependent on size and shape. To which extent can we still use the thermodynamic tools on nanoscale systems that apply to macroscale systems? When additional independent variables are needed in the Gibbs equation, from which pool do we draw them and how?

The free energies of a molecular fluid in a slit pore will be investigated at two conditions; at constant volume and constant normal pressure. In the first case, the system is in the canonical ensemble, and in the second case, the system is in the isobaric ensemble. In the thermodynamic limit, where the free energy is convex, we can transform the free energy from one ensemble to another using the Legendre transform. For small systems, this is not always the case. However, in the case of polymer stretching, we have recently found that the Legendre-Fenchel transform can be used \cite{bering2020legendre}. This experience has led us to wonder whether Legendre-Fenchel transforms can be used also for fluids confined to the slit pore, thereby motivating this paper. In the isochoric ensemble, the volume of the slit pore is controlled, and we will consider this as a simple model of a non-deformable porous medium. In the isobaric ensemble, the pore normal pressure is controlled, and the pore volume can fluctuate. We will consider this as a simple model of a deformable porous medium. 

The paper is outlined as follows. We give the theoretical background in section \ref{sec:theory}, including the Legendre-Fenchel transform in section \ref{sec:lf_transform}. We proceed to present the simulation technique in section \ref{sec:simulation} and show in section \ref{sec:results}, that the Helmholtz energy of an isochoric slit pore can be transformed into the Gibbs energy of an isobaric slit pore using the Legendre-Fenchel transform. The findings are discussed and perspectives are pointed out. In short, we shall see that systems that are small in Hill's sense have additional transitions than the bulk systems have. The system is more restricted when we control the fluid height than when we control the normal pressure.  

\subsection{System description.}
\begin{figure}
    \centering
    \includegraphics[width=\linewidth]{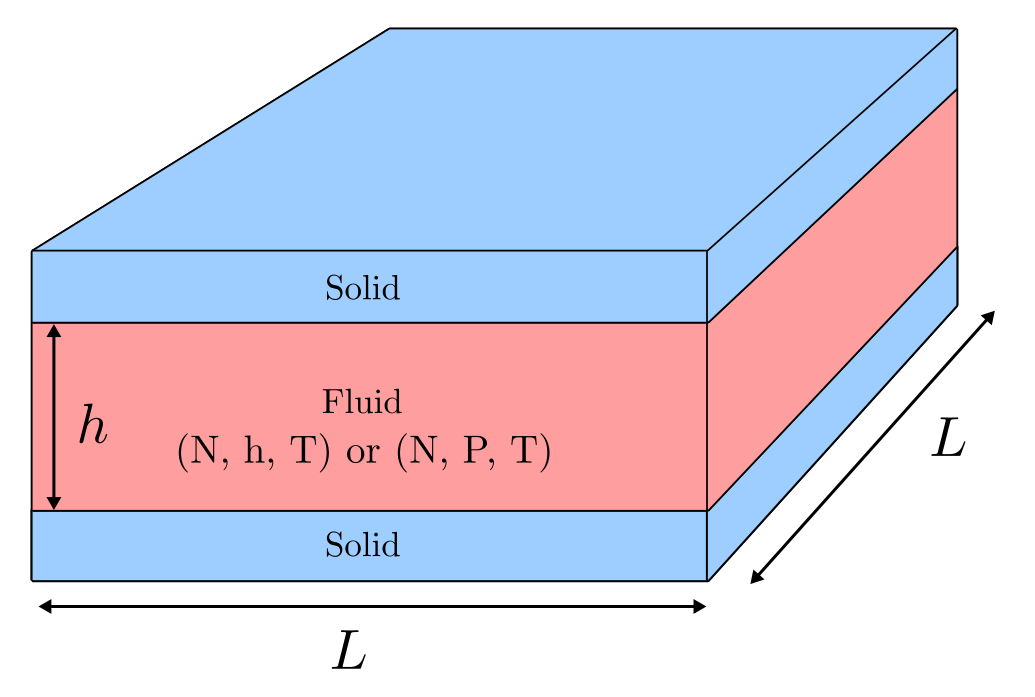}
    \caption{The fluid is placed between two parallel solid walls. The walls are separated by a distance $h$ (height) and the side lengths of the walls are $L$. In the canonical ensemble the height $h$ is controlled and the normal pressure $P$ fluctuates, while in the isobaric ensemble the normal pressure $P$ is controlled and the height $h$ fluctuates.}
    \label{fig:system}
\end{figure}

We investigate a single-phase fluid in a slit pore. This can be seen as a simple model suited to bring out the features described above. The system consists of a fluid placed between two parallel solid walls, see Fig. \ref{fig:system}. The system has periodic boundary conditions in the $y$- and $z$-directions. In the canonical ensemble, the walls do not move, while in the isobaric ensemble, the top wall can move in the $x$-direction and will act as a piston with controlled normal pressure on the fluid.

In the canonical ensemble the volume of the system is controlled, and the normal pressure $P$ fluctuates. In the isobaric ensemble, the normal pressure $P$ is controlled, and the volume $V=L^2h=Ah$ fluctuates, where $A=L^2$ is the fluid-solid surface area of one of the walls. The side lengths in the $y$- and $z$-directions are fixed equal to $L$, and it is only the distance between the walls $h$ (height) that fluctuates. The side lengths $L$ are much larger than the height such that the system may be considered to be independent of the surface area $A$.

\section{Theory}
\label{sec:theory}
In the canonical ensemble, the Helmholtz energy describes the maximum obtainable work of the system. The total differential of the Helmholtz energy is
\begin{equation}
    \d F(N,h,T) = -S\d T - P A \d h + \mu\d N,
\end{equation}
where $N$ is the number of fluid particles, $T$ is the temperature, $S$ is the entropy, $P$ is the normal pressure, $A$ is the fluid-solid surface area, and $\mu$ is the chemical potential. In the isobaric ensemble, it is the Gibbs energy that describes the maximum obtainable work of the system. The total differential of the Gibbs energy is
\begin{equation}
    \d G(N,P,T) = -S\d T + Ah\d P + \mu \d N.
\end{equation}
The normal pressure and height are defined in terms of the free energies as
\begin{equation}
    P \equiv -\frac{1}{A}\left(\frac{\partial F}{\partial h}\right)_{T, N}\quad\text{and}\quad h\equiv\frac{1}{A}\left(\frac{\partial G}{\partial P}\right)_{T, N}.
    \label{eq:pressure_def}
\end{equation}

The difference in the specific Helmholtz energy $\Delta f=F/M$, where $M$ is the total mass of the fluid, is calculated by integrating the mean normal pressure as a function of the volume in the canonical ensemble
\begin{equation}
    \Delta f = f(N,h,T)-f(N,h_0,T) = -\frac{A}{M}\int_{h_0}^{h} \langle P\rangle \d h',
    \label{eq:numerical}
\end{equation}
where $\langle P\rangle$ is the mean normal pressure. Similarly, the difference in the specific Gibbs energy $\Delta g=G/M$ is calculated by integrating the mean height as a function of the normal pressure in the isobaric ensemble
\begin{equation}
    \Delta g = g(N,P,T)-g(N,P_0,T) = \frac{A}{M}\int_{P}^{P_0} \langle h\rangle\d P',
\end{equation}
where $\langle h\rangle$ is the mean height, and $f_0 = f(N,h_0,T)$ and $g_0 = g(N,P_0,T)$ are reference states at a large height $h_0$ and low pressure $P_0$. The difference in the specific entropy of the system in the isochoric ensemble is
\begin{equation}
    \Delta s = \frac{1}{T}(\Delta u-\Delta f),
\end{equation}
where $\Delta u$ is the change in the specific internal energy of the system. The changes in specific entropy and internal energy are relative to a system at height $h_0$.  

\subsection{The Legendre-Fenchel transform.}
\label{sec:lf_transform}

In the isobaric ensemble, the control variables are temperature $T$, normal pressure $P$, and the number of fluid particles $N$. The normal pressure is equal to the absolute force acting on the walls divided by the surface area $A$. This is equal to the normal component of the mechanical pressure tensor of the fluid \cite{galteland2021nanothermodynamic}. The isobaric partition function can be obtained by a Laplace transform of the canonical partition function
\begin{equation}
    \mathcal{Z}(N,P,T) = \beta P A\int_0^\infty Z(N,h,T)\exp(-\beta P Ah)\d h,
    \label{eq:isothermal_isobaric_partition_function}
\end{equation}
where $Z(N,h,T)$ is the canonical partition function and $\beta=(k_\text{B}T)^{-1}$ where $k_\text{B}$ is the Boltzmann constant. The Helmholtz and Gibbs energies are given by the corresponding partition functions
\begin{equation}
    F(N,h,T) = -k_\text{B}T\ln Z(N,h,T)
\end{equation}
and
\begin{equation}
    G(N,P,T) = -k_\text{B} T\ln \mathcal{Z}(N,P,T),
\end{equation}
respectively. From the three equations above it follows that the Gibbs energy can be obtained from the Helmholtz energy,
\begin{equation}
  \begin{split}
    &\exp[-\beta G(N,P,T)] = \\
    &\beta P A\int_0^\infty\exp[-\beta (F(N,h,T)+P Ah)]\d h.
  \end{split}
    \label{eq:transform}
\end{equation}
For sufficiently high surface number densities $\Gamma=N/A$, the system is large and the Helmholtz energy $F$ is a differentiable and convex function of the height, and the above expression reduces to the Legendre transform of the Helmholtz energy to the Gibbs energy,
\begin{equation}
    G_\text{L}(N,P,T) = G(N,h,T) + PhA.
\end{equation}
For low surface number densities $\Gamma$, the system is small, and the Helmholtz energy is non-convex and the integral in equation \ref{eq:transform} does not reduce to the Legendre transform. If the distribution of the normal pressure is sharply peaked it can however be calculated with a saddlepoint approximation \cite{bering2020legendre,touchette2005legendre, rockafellar2015convex},
\begin{equation}
    G_\text{LF}(N, P, T) = \min_{h}(F(N,h,T)+P hA).
    \label{eq:legendre_fenchel}
\end{equation}
This is the Legendre-Fenchel (LF) transform of the Helmholtz energy $F$ to the Gibbs energy $G_\text{LF}$. The LF transform returns only convex functions. If we apply it again,
\begin{equation}
    F^{**} = \max_p(G_\text{LF}+PhA) = G_\text{LF}+PhA,
\end{equation}
we obtain the convex envelope of the Helmholtz energy $F^{**}$. The convex envelope is the largest function satisfying $F^{**} \leq F$, which is only equal to the original Helmholtz energy $F$ if it is a convex function. In other words, the LF transform is not self-inverse if the function is non-convex \cite{touchette2005legendre, rockafellar2015convex}. The LF transform can be defined as either the maximum or minimum. Since $G_\text{LF}$ must be convex, we can also obtain $F^{**}$ from a Legendre transform of $G_\text{LF}$. 

The Maxwell construction of equal areas for liquid-vapor coexistence is equivalent to the convex envelope of the Helmholtz energy $F^{**}$. The equal area rule states that for a liquid-vapor coexistence the system follows a constant pressure $P_\text{eq}$ from volume $V_\text{l}$ to $V_\text{g}$ when the system evaporates, and conversely for condensation. The two volumes $V_\text{l}$ and $V_\text{g}$ at the pressure $P_\text{eq}$ are the binodal points of the pressure-volume curve. The equal area rule states
\begin{equation}
    \label{eq:equal_area}
    \int_{V_\text{l}}^{V_\text{g}} P(V)\d V = P_\text{eq}(V_\text{l}-V_\text{g})
\end{equation}
where $P(V)$ is a cubic equation of state, for example, the van der Waals equation, below the critical point. This corresponds to finding the double tangent line of the non-convex Helmholtz energy. The double tangent line is a tangent of the function at two different points. A double tangent line does not exist for convex functions. The double tangent line is exactly the convex envelope, as it is the largest function that satisfies $F^{**} \leq F$.

\section{Simulation details}
\label{sec:simulation}
A fluid between two parallel solid walls in the canonical and the isobaric ensemble was investigated with molecular dynamics simulations using LAMMPS \cite{plimpton1995fast,thompson2021lammps}. The temperature of the fluid was controlled using the Nosé-Hoover thermostat to $T = 2.26T_c=2\varepsilon/k_\text{B}$, where $T_c$ is the critical temperature of the bulk fluid \cite{hafskjold2019thermodynamic}. 

\begin{figure}
    \centering
    \includegraphics[width=\linewidth]{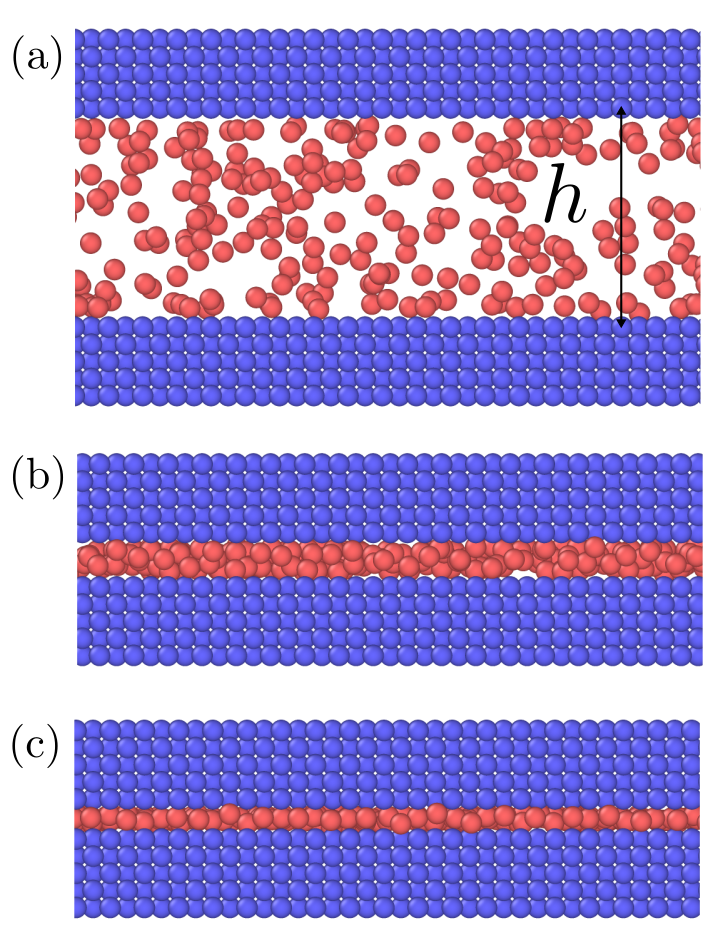}
    \caption{A rendering of simulations in the isobaric ensemble with surface number density $\Gamma=0.81$. The top solid wall is the free to move in the normal direction and acts as a piston on the fluid with a controlled normal pressure $P$. The height $h$ is shown in \textbf{(a)} and is defined as the minimum distance between the center of a solid particle in the bottom solid wall and the center of a solid particle in the piston. The normal pressure and mean height were equal to \textbf{(a)} $P=0.164\varepsilon/\sigma^3$ and $\langle h\rangle = (10.2\pm0.09)\sigma$, \textbf{(b)} $P=2.88\varepsilon/\sigma^3$ and $\langle h\rangle = (2.530\pm0.009)\sigma$ and, \textbf{(c)} $P=2.89\varepsilon/\sigma^3$ and $\langle h\rangle = (1.865\pm0.003)\sigma$.}
    \label{fig:simulation}
\end{figure}

The walls were made up of solid particles in a face-centered cubic lattice with a number density $n_s^* = 1/\sigma^3$ corresponding to a lattice constant $a=2^{2/3}\sigma$, where $\sigma$ is the fluid particle diameter. Each wall had $N^p=5\times 10^4$ solid particles and side lengths $L=100a\approx159\sigma$. Each solid particle had a mass equal to the fluid particle mass, implying that the mass of the top solid wall (piston) was $5\times 10^4$ times greater than a fluid particle. The mass of each particle were equal to $m$, which in reduced units is equal to one. The total mass of the fluid particles were $M=Nm$. In the canonical ensemble, the solid particles were fixed in space and could not move. In the isobaric ensemble, the solid particles in the lower wall were fixed in space, while the solid particles in the piston were free to move as a single rigid body in the $x$-direction. The piston could not rotate or move in the $y$- or $z$-directions. The fluid particles were placed between the walls. The system is visualized in Fig. \ref{fig:simulation} using OVITO \cite{stukowski2009visualization}. The fluid particles are drawn in red and the solid particles are drawn in blue.

The fluid-fluid and fluid-solid particles interacted with the Lennard-Jones/spline potential \cite{hafskjold2019thermodynamic},
\begin{equation}
	u^\text{LJ/s}(r)= 
	\begin{cases} 
	    4\varepsilon\left[\left(\frac{\sigma}{r}\right)^{12}-\left(\frac{\sigma}{r}\right)^{6}\right] &\text{if } r<r_{s},\\
	    a(r-r_{c})^{2}+b(r-r_{c})^{3} &\text{if } r_{s}<r<r_{c},\\
	    0 &\text{else},
	\end{cases}
	\label{eq:ljs}
\end{equation}
where $r=|\pmb{r}_j-\pmb{r}_i|$ is the distance between particle $i$ and $j$, $\sigma$ is the particle diameter, and $\varepsilon$ is the well-depth of the interactions. The parameters $r_s=(26/7)^{(1/6)}\sigma\approx1.24\sigma$, $a=-24192/3211(\varepsilon/r_s^2)$, $b=-387072/61009(\varepsilon/r_s^3)$ and $r_c=67/48r_s\approx1.74\sigma$ were set such that the potential energy and the force were continuous at $r_s$ and $r_c$. An advantage of the Lennard-Jones/spline potential is that the cut-off is much shorter than the regular Lennard-Jones potential with a typical cut-off at $2.5\sigma$, which considerably decreases the computation time. See for more details on the properties of the Lennard-Jones/spline potential the work of Hafskjold \textit{et al} \cite{hafskjold2019thermodynamic} and Kristiansen \cite{kristiansen2020transport}. The fluid-fluid and fluid-solid interactions were equal, while the solid-solid interaction was zero. The size of the timesteps was set equal to $\delta t = 0.002$. All units are in reduced Lennard-Jones units, meaning that they are reduced with the mass, particle diameter $\sigma$, the minimum of the interaction potential $\varepsilon$, and the Boltzmann constant $k_\text{B}$. 

The system was initialized by creating one slab of fluid particles between two slabs of solid particles, all in a face-centered cubic lattice. The fluid particles were initialized with a velocity such that the temperature was equal to $T = 2\varepsilon/k_\text{B}$, and they were free to move for $10^4$ timesteps to melt the face-centered cubic lattice of the fluid. In the canonical ensemble, the piston was moved with a constant velocity for $10^5$ timesteps to reach the desired height $h$, from here on defined as the minimum distance between the center of a solid particle in the bottom solid wall and the center of a solid particle in the piston. The controlled height was in the range $h\in[1.7, 110]\sigma$. Then the position of the piston was fixed, and the system was run for $10^6$ timesteps to calculate the mean normal pressure, $\langle P\rangle$. 

The mean normal pressure was calculated as the arithmetic mean of the instantaneous forces of the particles in the piston at time $t$ at time intervals $\Delta t = 0.2$
\begin{equation}
    \langle P\rangle = \frac{1}{An}\sum_{t=0}^n\sum_{i=1}^{N^p}f_{x,i,t},
\end{equation}
where $f_{x,i,t}$ is the force in the $x$-direction acting on the solid particle $i$ in the piston at time $t$, $N^p$ is the number of solid particles, and $n=10^6(\delta t/\Delta t)=10^4$ is the number of samples where $\delta t$ is the size of the timestep. Alternatively, the normal pressure could be calculated as the normal component of the mechanical pressure tensor in the fluid \cite{galteland2021nanothermodynamic}. This has been done as a consistency check. 

In the isobaric ensemble, an external force was added to the piston after melting the face-centered cubic lattice of the fluid. The simulations with controlled normal pressures were run sequentially from low to high normal pressure to obtain the compression curve, and from high to low normal pressure to obtain the expansion curve. This was done to reach all available states of the system in the isobaric ensemble. The controlled normal pressures were in the range $P\in[0.0045, 25]\varepsilon/\sigma^3$. Each of the simulations was run for $10^6$ timesteps to calculate the mean height $\langle h\rangle$.

An external force $f_x$ in the $x$-direction was applied to each solid particle in the piston,
\begin{equation}
    f_x = \frac{P A}{N^p}
\end{equation}
where $N^p$ is the number of particles in the piston. The mean height $\langle h \rangle$ was calculated as the arithmetic mean of the instantaneous height at time $t$ at time intervals of $\Delta t = 0.2$
\begin{equation}
    \langle h \rangle = \frac{1}{n}\sum_t^n h_t.
\end{equation}

The mean of the specific internal energy was calculated as the mean potential energy plus the mean kinetic energy of the fluid particles,
\begin{equation}
    \langle u\rangle = \frac{1}{M}\left(\sum_{i=1}^N\sum_{j>i}^N u^\text{LJ/s}(r) + \frac{1}{2}\sum_{i=1}^Nm(\pmb{v}_i\cdot\pmb{v}_i)\right)
\end{equation}
where $\pmb{v}_i$ is the velocity of particle $i$, $m$ is the mass of each fluid particle and $M=Nm$ is the total mass of the fluid particles. The specific internal energy was used together with the specific Helmholtz energy to calculate the specific entropy. The reference states $f_0$, $u_0$, and $s_0$ were calculated at height $h_0=110\sigma_0$, and $g_0$ was calculated at pressure $P_0$ corresponding to a mean height $\langle h_0\rangle=110\sigma_0$. 

\section{Results and discussion}
\label{sec:results}

\begin{figure}
    \centering
    \includegraphics[width=\linewidth]{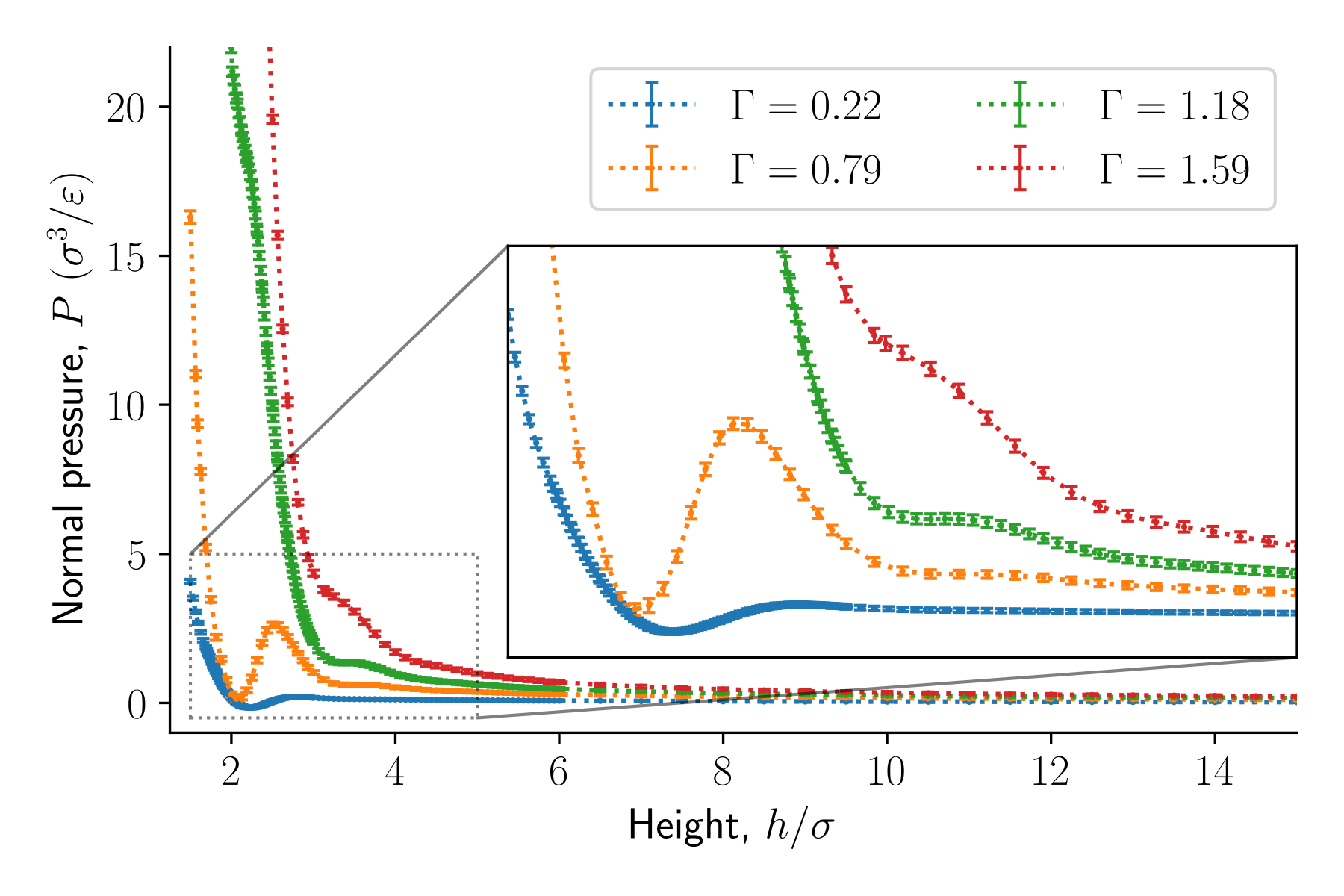}
    \caption{Normal pressure as a function of the height in isochoric conditions for varying surface number densities. The insert is an enlargement of the region where structural transitions occur.}
    \label{fig:pressure_height_combined}
\end{figure}

\begin{figure}
    \centering
    \includegraphics[width=\linewidth]{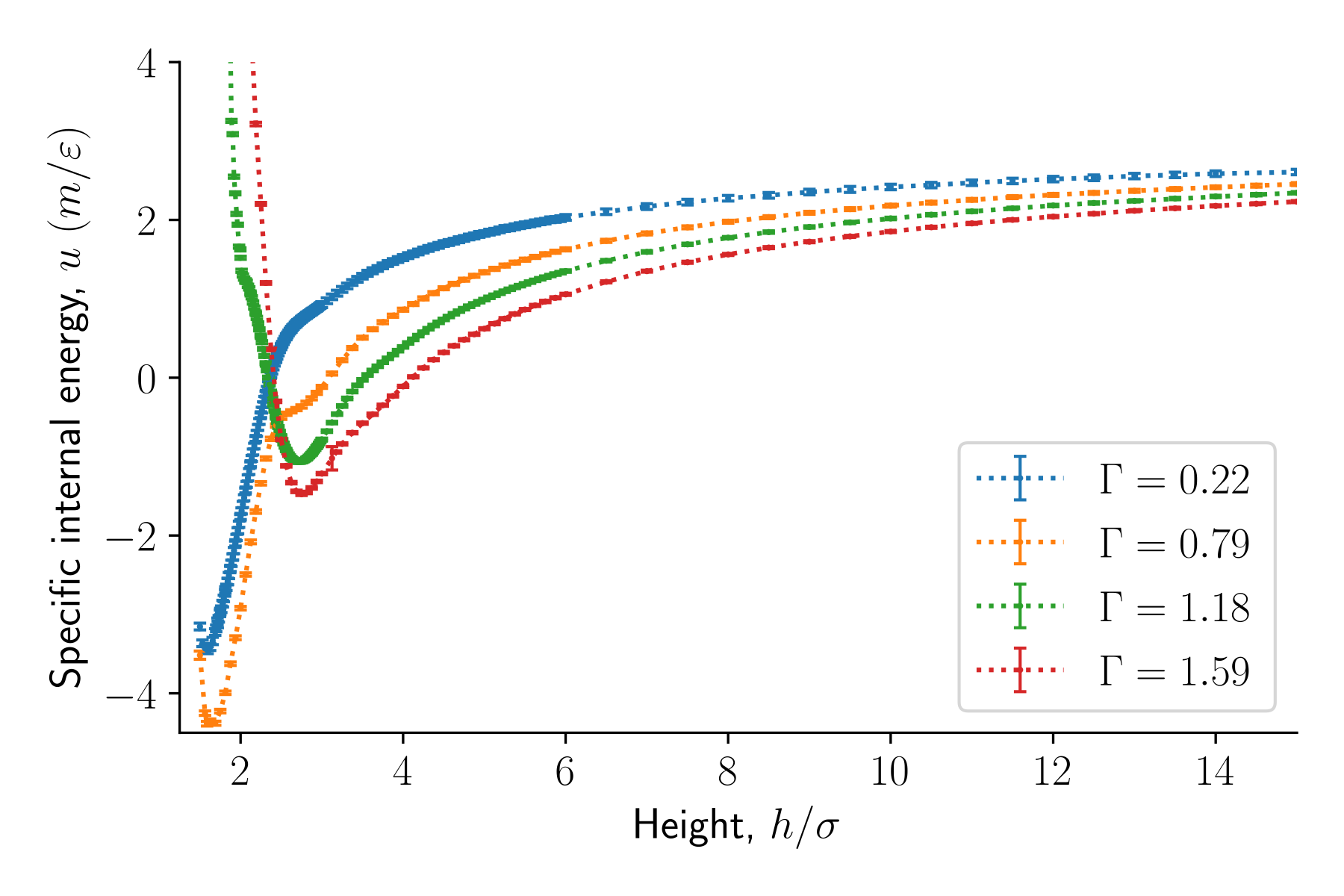}
    \caption{Specific internal energy as a function of the height in isochoric conditions for varying surface number densities.}
    \label{fig:internal_energy}
\end{figure}

\begin{figure}
    \centering
    \includegraphics[width=\linewidth]{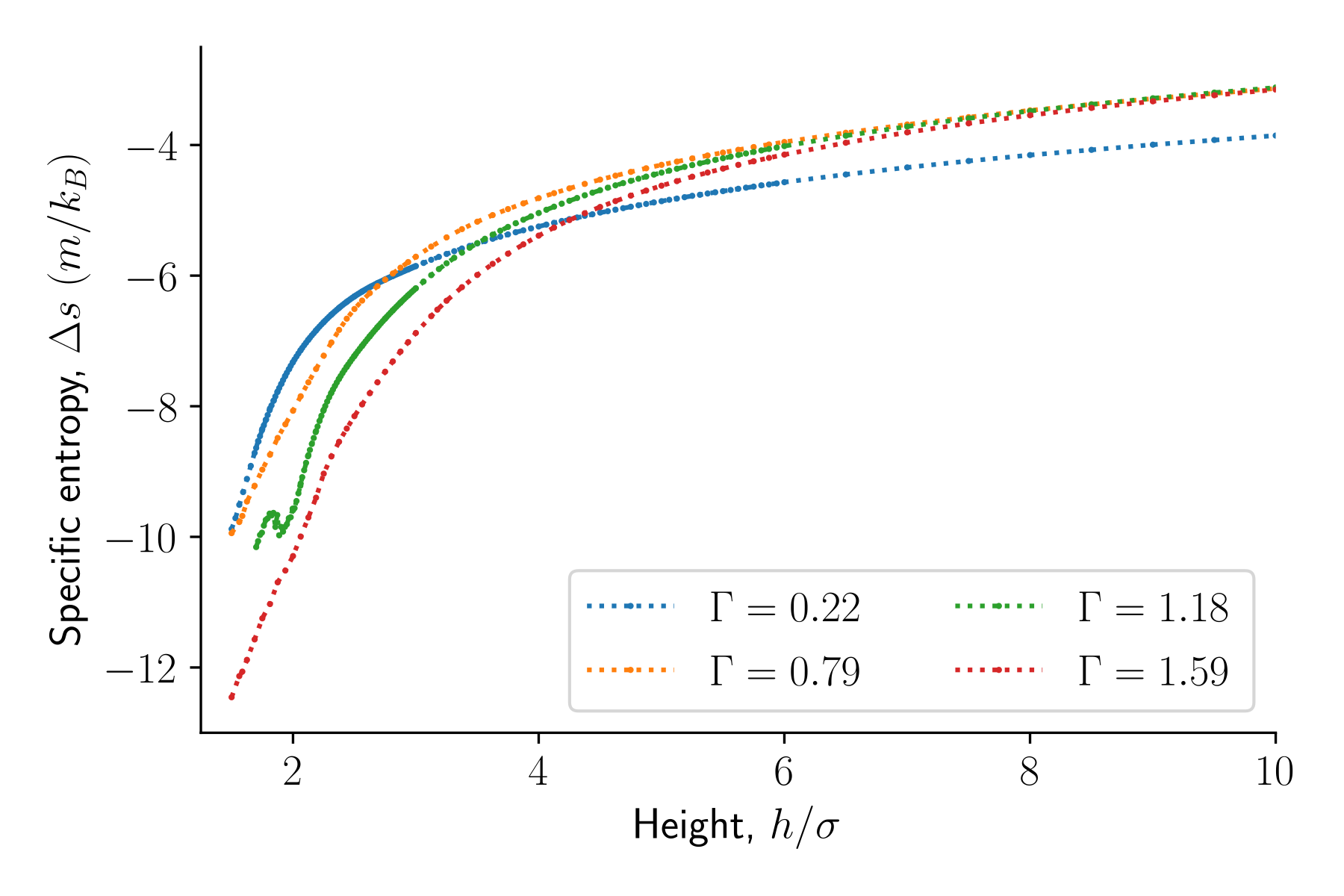}
    \caption{Difference in specific entropy as a function of the height in isochoric conditions for varying surface number densities.}
    \label{fig:entropy}
\end{figure}

\begin{figure}
    \centering
    \includegraphics[width=\linewidth]{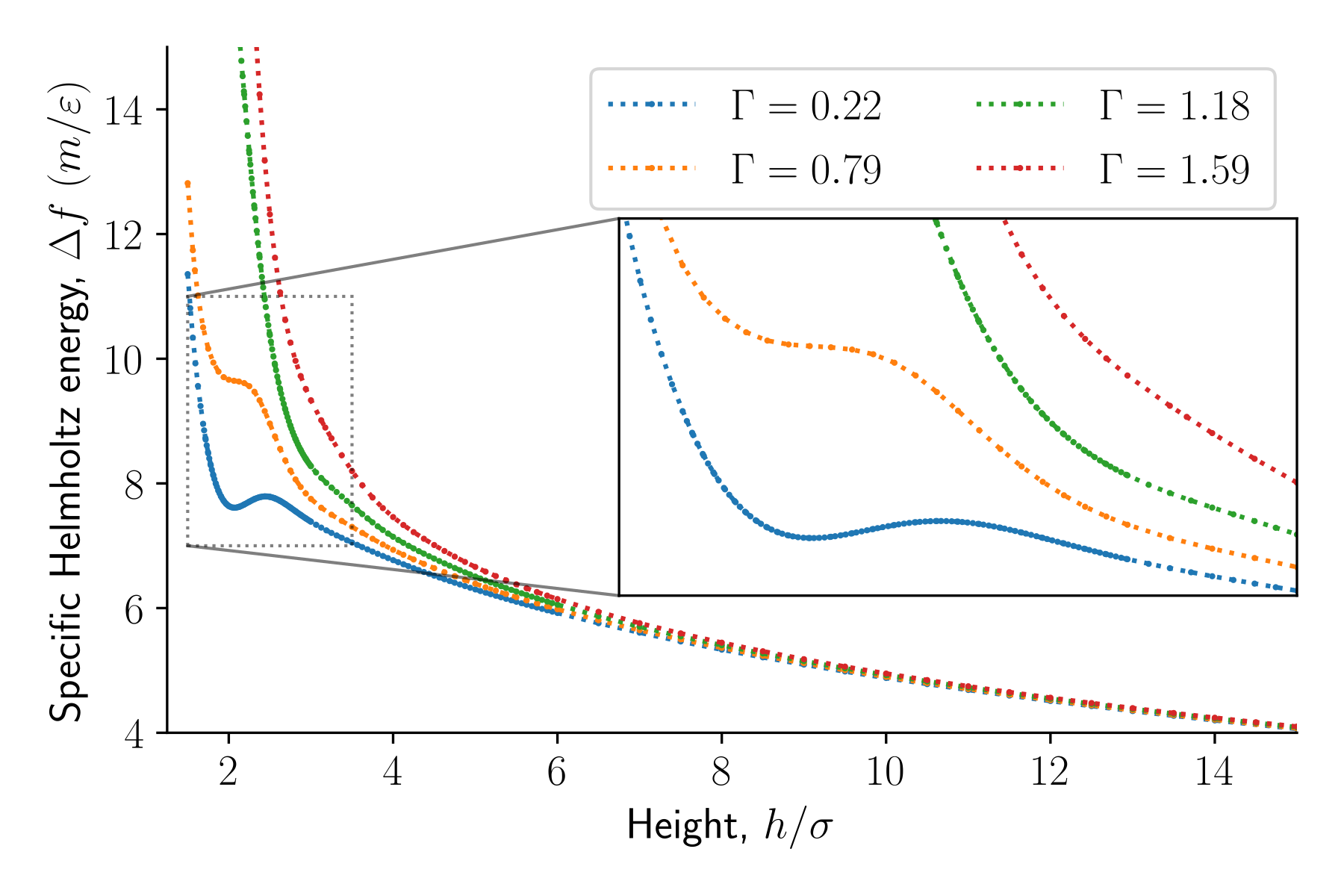}
    \caption{Difference in specific Helmholtz energy as a function of height for isochoric conditions for varying surface number densities.}
    \label{fig:helmholtz_energy}
\end{figure}

\begin{figure}
    \centering
    \includegraphics[width=\linewidth]{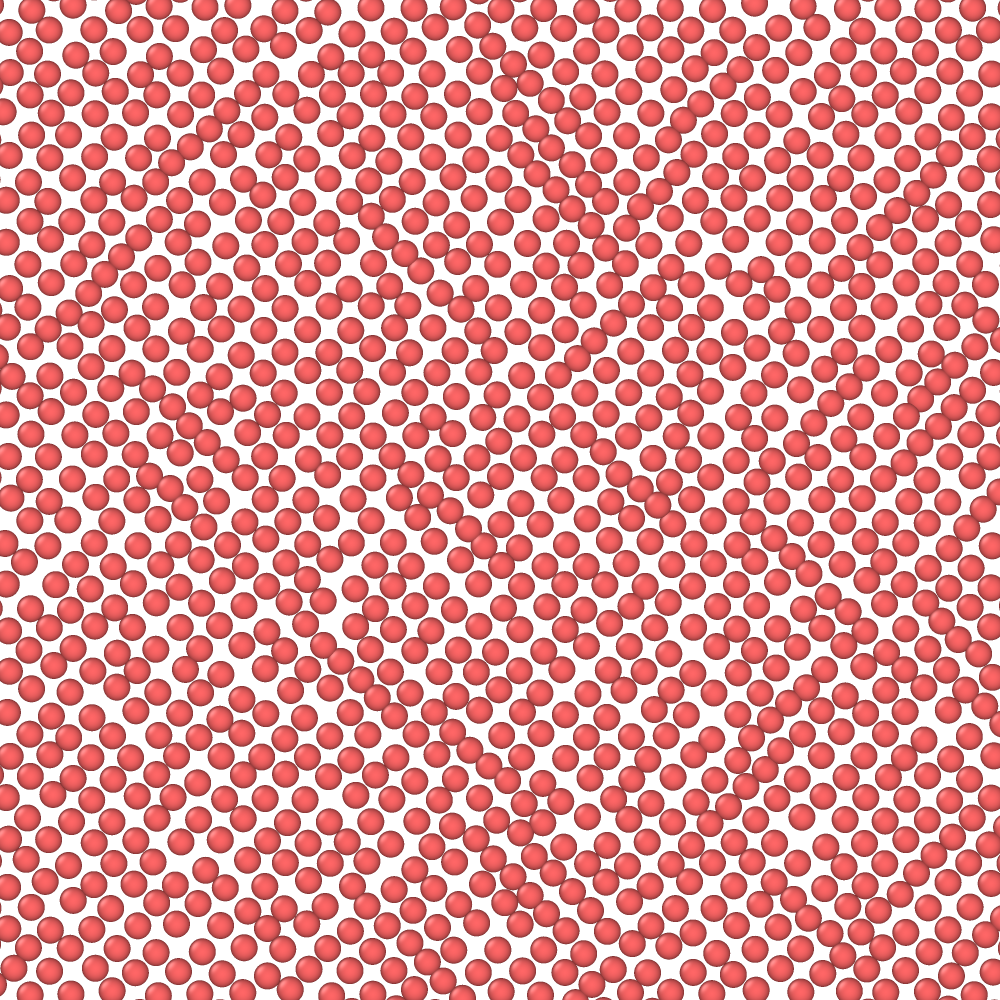}
    \caption{Top down view of fluid particles in isochoric conditions which has formed a hexagonal monolayer at height $h=2^{2/3}\sigma\approx1.59\sigma$. The solid particles are not shown. The mean normal pressure is $\langle P\rangle = (9.4\pm0.1)\varepsilon/\sigma^3$, specific internal energy $\langle u \rangle =(-4.39\pm0.03)\varepsilon/m$, difference in the specific entropy $\langle \Delta s\rangle = -9.7 k_\text{B}/m$, and difference in specific Helmholtz energy is $\langle \Delta f\rangle = 11.4\varepsilon/m$ \cite{stukowski2009visualization}.}
    \label{fig:monolayer}
\end{figure}

The results are illustrated in Figs. \ref{fig:pressure_height_combined} to \ref{fig:free_energy_height}. Fig. \ref{fig:pressure_height_combined} shows the normal pressure-height relationship in isochoric conditions for various surface number densities. Figs. \ref{fig:internal_energy} to \ref{fig:helmholtz_energy}, give the specific internal energy, difference in specific entropy, and difference in specific Helmholtz energy, respectively, all properties as a function of a controlled height. The difference in specific entropy and Helmholtz energy are given as the difference relative to the reference state at $h_0=110\sigma$. Fig. \ref{fig:monolayer} is a visualisation of the monolayer in isochoric conditions when close to hexagonal packing appears. 

The impact of the choice of environmental control variables on the normal pressure-height relationship, the basis of the findings reported, is illustrated in Fig. \ref{fig:pressure_height}. This provides a basis for examination of Legendre and the Legendre-Fenchel transforms, see Figs. \ref{fig:free_energy_pressure} and \ref{fig:free_energy_height}. The various results will now be explained and discussed. 

\subsection{The normal pressure, specific internal energy, entropy, and Helmholtz energy as a function of height.}

The normal pressure-height relationship as a function of height and for various number densities was presented in Fig. \ref{fig:pressure_height_combined}. The thermodynamic limit behavior is seen for large heights or large surface number densities. The large system has a differentiable convex Helmholtz energy, approximately when the height $h>3\sigma$ and when the surface number densities are $\Gamma\geq1.18$. The typical small system behaviour appears for heights $h<3\sigma$ and surface densities $\Gamma < 1.18$.

For smaller heights or surface number densities, the normal pressure goes through a local minimum and maximum as the height changes. This implies that the Helmholtz energy is non-convex. The specific internal energy has also a minimum, see Fig. \ref{fig:internal_energy}. The entropy is monotonically increasing, except for the case $\Gamma = 1.18$ which has a local minimum, see Fig. \ref{fig:entropy}. The Helmholtz energy in Fig. \ref{fig:helmholtz_energy} captures the trade-off between the internal energy and the entropy. The Helmholtz energy as a function of height is non-convex for surface densities $\Gamma<1.18$, which entails that the Legendre transform can not be applied. The insert in Fig. \ref{fig:helmholtz_energy} magnifies the interesting region. The emerging structures are stabilized by the ability of the system to go to lower energy and higher entropy. The curves indicate a smooth structural transition.

\subsection{A small system phase transition.}

Figs. \ref{fig:pressure_height_combined} to \ref{fig:helmholtz_energy} showed a family of curves that represent the small system in a transition region for small values of $h$. This transition is special for a small thermodynamic system, it disappears when surface number density increases. How can we understand better the behavior of the particles in this region?

For the system to change from a fluid to a close-packing structure (face-centered cubic or hexagonal close-packed) without defects, the surface number density $\Gamma$ must be
\begin{equation}
    \Gamma = \frac{k}{r^2},
\end{equation}
where $k$ is the number of layers (for a monolayer $k=1$) and $r$ is the distance between the fluid particles. The total potential energy is at a minimum at the distance $r=2^{1/6}\sigma\approx1.12\sigma$, which is where the Lennard-Jones/spline interaction potential is at a minimum. The interaction potential is short-ranged, it is zero for distances larger than the cut-off at $r_c\approx1.74\sigma$. The surface number density for a monolayer ($k=1$) is $\Gamma\approx0.79$, and for a double layer ($k=2$) it is $\Gamma\approx1.59$. An odd number of layers can complete the face-centered cubic lattice of walls without faults, while an even number of layers must have an odd number of stacking faults. 

In the extreme case of a monolayer, we assume that the fluid is packed in such a way that each fluid particle lies on average at a distance $r_0 = 2^{1/6}\sigma$ away from eight solid particles and four fluid particles. In addition, there are two solid particle neighbours and four fluid particle neighbours each at a distance $r_a = 2^{2/3}\sigma$. All other particles lie beyond the cutoff distance $r_c$ and do not contribute to the potential energy.

The minimum specific potential energy of a monolayer is then
\begin{equation}
e_{p,0} = 10u^\text{LJ/s}(r_0)+4u^\text{LJ/s}(r_a)\approx -10.39\varepsilon/m,
\end{equation}
where $m$ is the particle mass. Taking into account thermal fluctuations, it is
\begin{equation}
\langle e_p\rangle = e_{p,0} + a\langle r^2\rangle
\end{equation}
where $a$ is a constant. Assuming no defects, the average deviation from the average position can be set to zero. The quadratic term represents the lowest order correction to the mean specific potential energy due to thermal fluctuations making the fluid particle spend on average more time away from the average position. By the equipartition theorem, we then obtain the mean potential energy of a fluid particle
\begin{equation}
\langle e_p\rangle = e_{p,0} + \frac{3}{2}k_\text{B}T\approx -7.39\varepsilon/m.
\end{equation}
The expected minimum mean specific internal energy is the sum of the mean specific potential and kinetic energies
\begin{equation}
\langle u\rangle_\text{min} = \langle e_p\rangle + \langle e_k\rangle \approx -4.39\varepsilon/m
\end{equation}
where we have used that the mean specific kinetic energy is $\langle e_k\rangle = 3k_\text{B}T/(2m) = 3\varepsilon/m$ at a temperature $T=2\varepsilon/k_\text{B}$.

The observed minimum of the specific internal energy for $\Gamma=0.79$ is indeed $\langle u\rangle =(-4.39\pm0.03)\varepsilon/m$ at $h=2^{2/3}\sigma\approx1.59\sigma$, see Fig. \ref{fig:internal_energy}. The simulated structure is visualised in Fig. \ref{fig:monolayer}. The figure shows that the fluid particles have formed a hexagonal monolayer layer. The mean normal pressure is $\langle P\rangle = (9.4\pm0.1)\varepsilon/\sigma^3$, difference in the specific entropy $\langle \Delta s\rangle = -9.7 k_\text{B}/m$, and difference in specific Helmholtz energy is $\langle \Delta f\rangle = 11.4\varepsilon/m$. The specific internal energy is at a minimum at this height, however, the Helmholtz energy is not. 

A transition from a fluid to a close-packed structure under stress does not imply a first-order phase transition, a continuous transition from a fluid to a close-packed structure packing may occur. In other words, the free energy can be smooth and continuous during the transition.

Consider for comparison the familiar pressure-volume isotherms of cubic equations of state, for example, the van der Waals equation, for temperatures below the critical point. The Helmholtz energy is a non-convex function of volume. The binodal curve intersects the pressure-volume isotherms at two points $a=(p_\text{eq}, V_\text{l})$ and $b=(p_\text{eq}, V_\text{g})$ for $T<T_c$. The part of the isotherm between point $a$ and $b$ is known as the van der Waals loop. The Helmholtz energy is a non-convex function in this region. The line between the points $a$ and $b$ corresponds to the Maxwell construction of equal areas. See equation \ref{eq:equal_area}. The double Legendre-Fenchel transform of the specific Helmholtz energy $f$ gives its convex envelope $f^{**}$, which corresponds to the Maxwell construction of equal areas. The spinodal curve intersects the isotherm at the local minimum and maximum, the spinodal region is a subset of the binodal region. The binodal region is metastable, while the spinodal region is unstable. Experimentally it is observed that fluids do not necessarily follow the van der Waals loop, but rather the straight line connecting points $a$ and $b$. This is a first-order phase transition, as the pressure is non-smooth at the points $a$ and $b$. The system is free to decompose in the binodal and spinodal region, which is energetically more favorable. During liquid-vapor phase decomposition the pressure is constant and equal to $P_\text{eq}$. The constant pressure corresponds to the double tangent line in the free energy. This double tangent line is the largest convex curve that satisfies $f^{**} \leq f$, which is exactly its convex envelope.

\begin{figure}
    \centering
    \includegraphics[width=\linewidth]{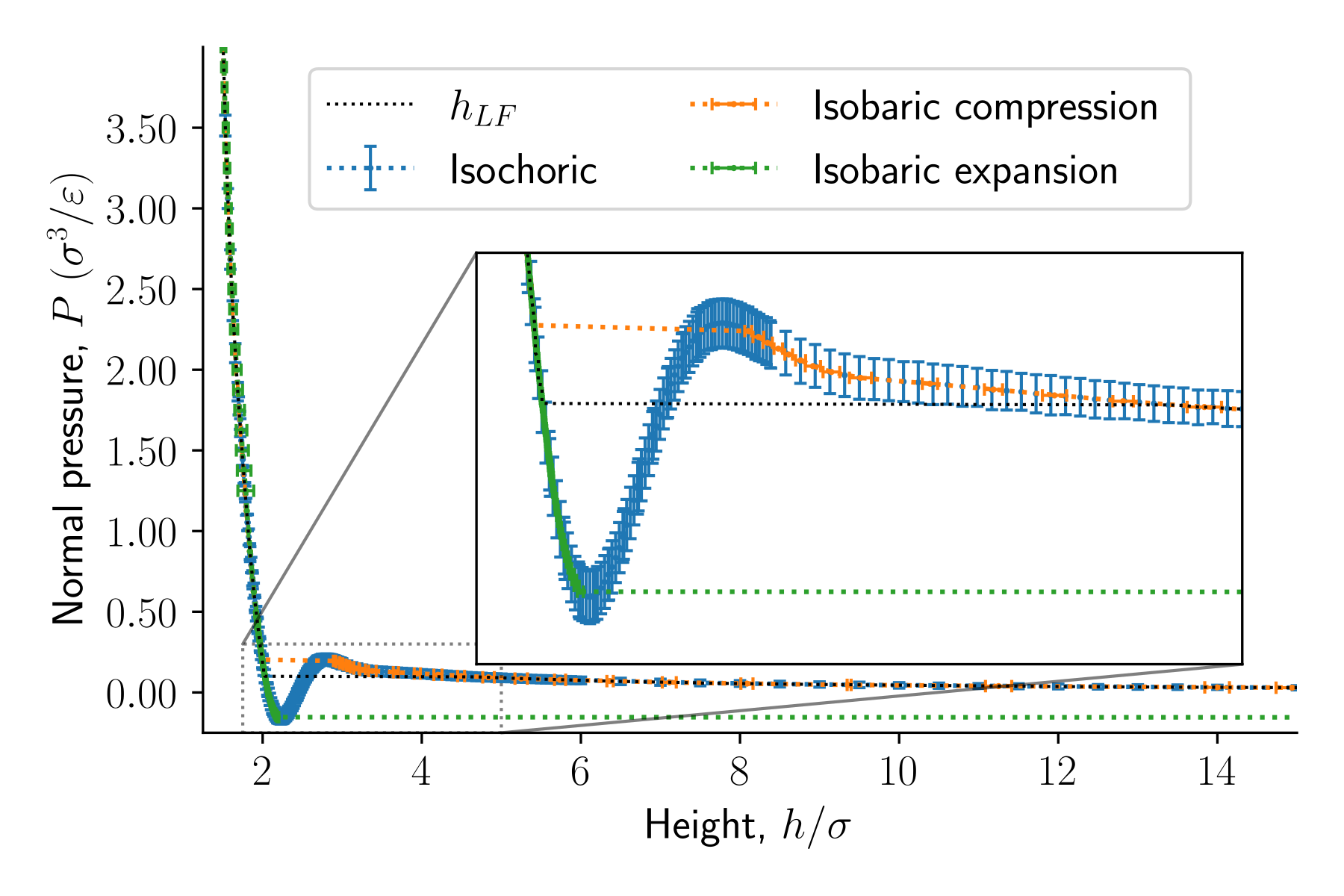}
    \includegraphics[width=\linewidth]{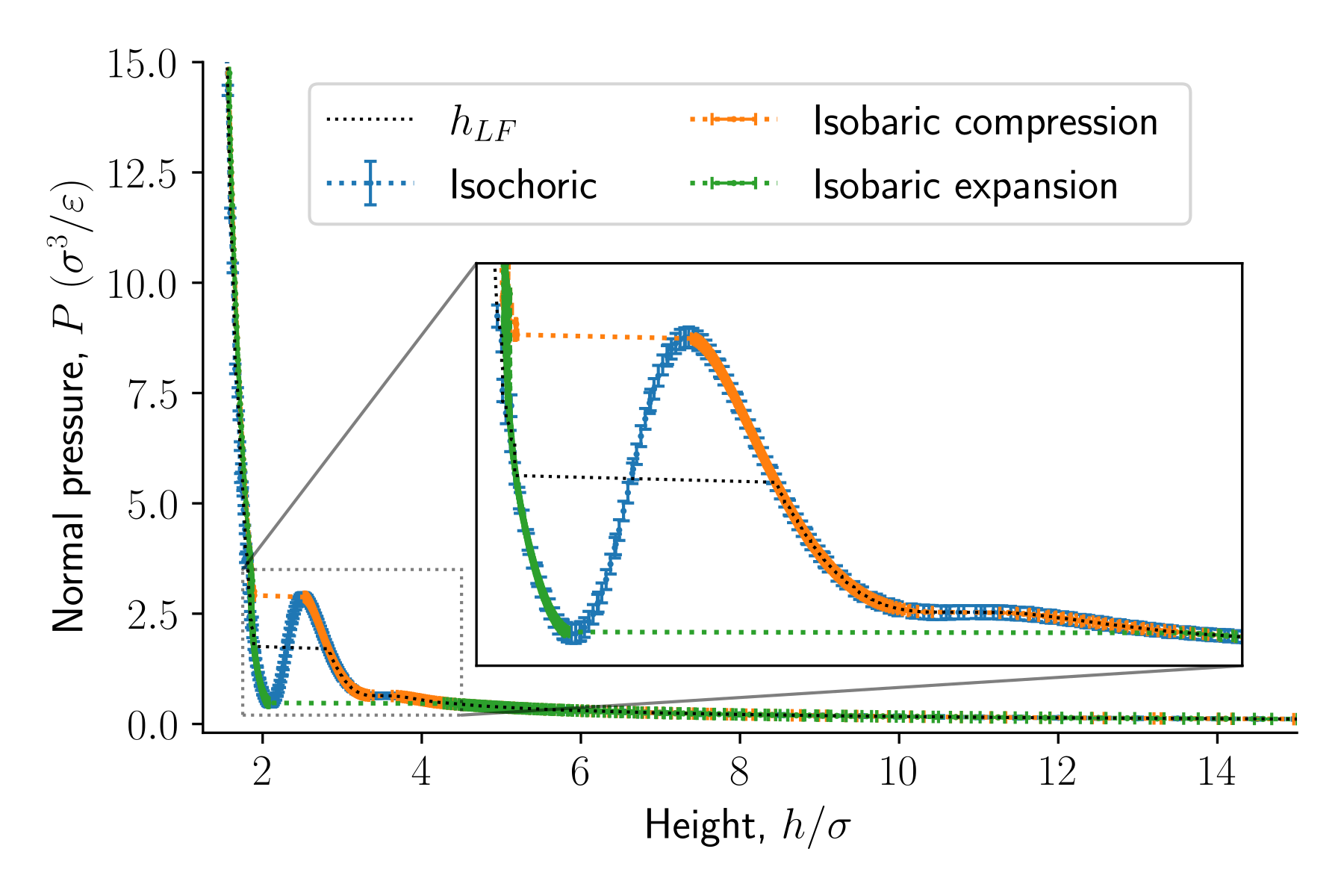}
    \includegraphics[width=\linewidth]{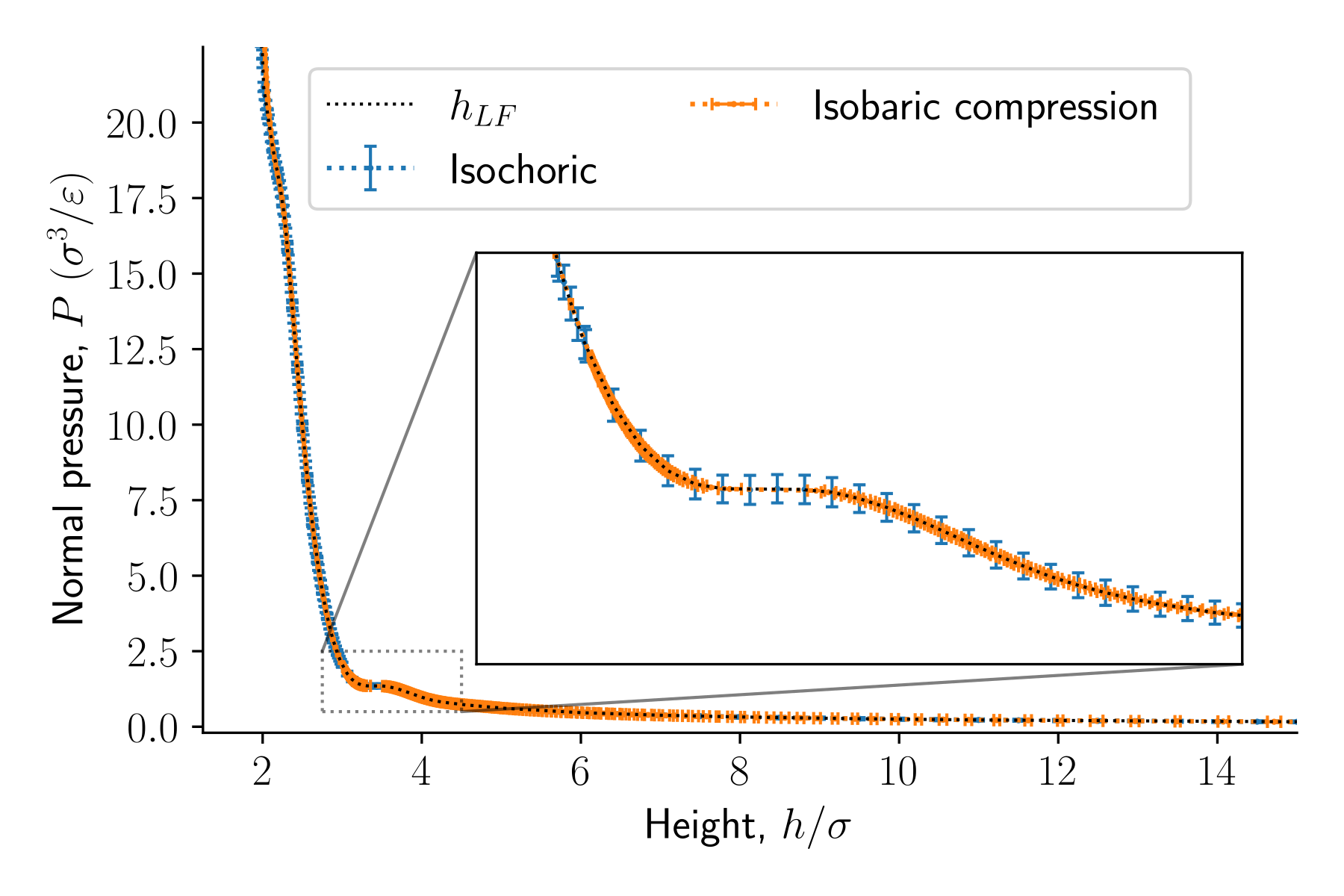}
    \caption{Normal pressure as a function of the height for isochoric and for isobaric conditions. The surface number density in each figure is $\Gamma=0.22$, $\Gamma=0.81$, and $\Gamma=1.18$ from top to bottom. The top and middle figures shows the isobaric expansion and compression curves.}
    \label{fig:pressure_height}
\end{figure}

The states in the binodal and spinodal regions can be stable due to the restrictions that the confinement imposes on the system. The coexistence of fluid and close-packed structures is not possible in these simulations, as this would imply that there would be regions with differing heights. This could be possible if the walls were free to rotate or deform, however, the system is restricted such that the height is everywhere the same. In isochoric conditions, there is a smooth transition from a fluid to a close-packed structure. In isobaric conditions, the system is less restricted, and the system undergoes a first-order phase transition when it enters the spinodal region, see Figs. \ref{fig:pressure_height} (center). This is because the spinodal region is unstable. The set of control variables provide different stable states with their different restrictions on the system. 

\begin{figure}
    \centering
    \includegraphics[width=\linewidth]{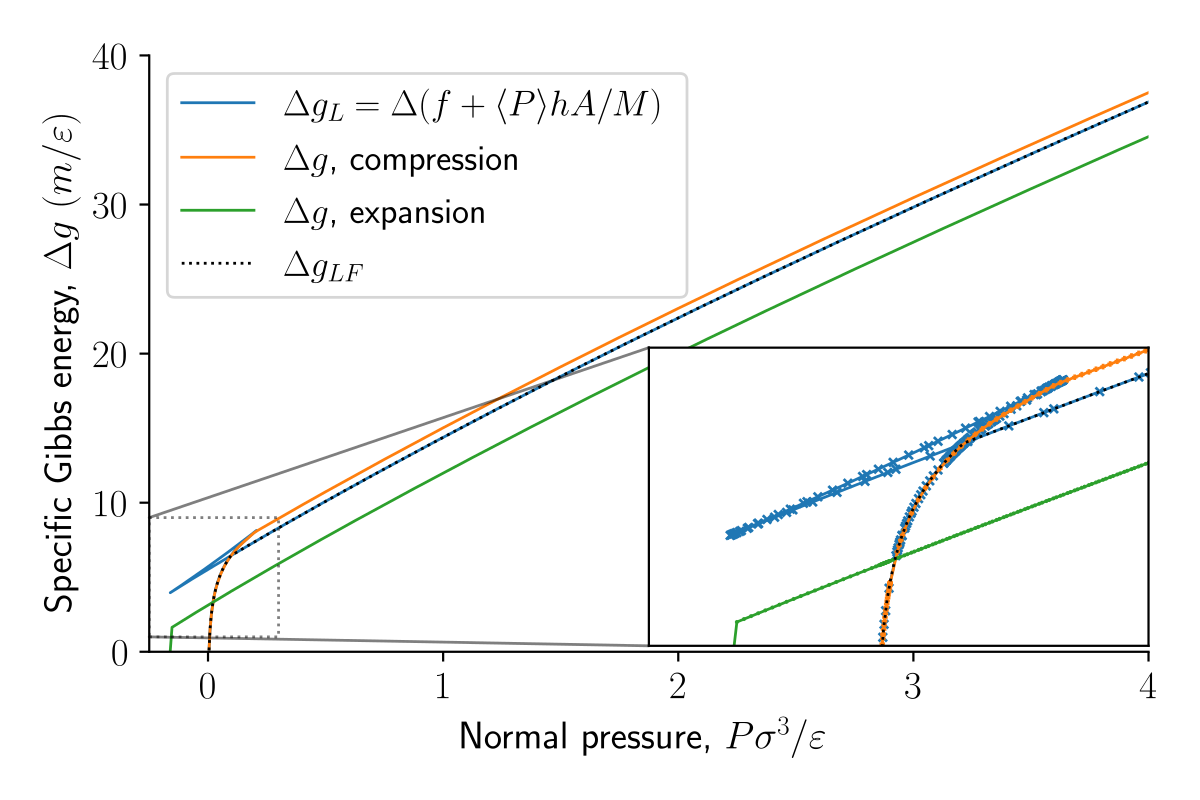}
    \includegraphics[width=\linewidth]{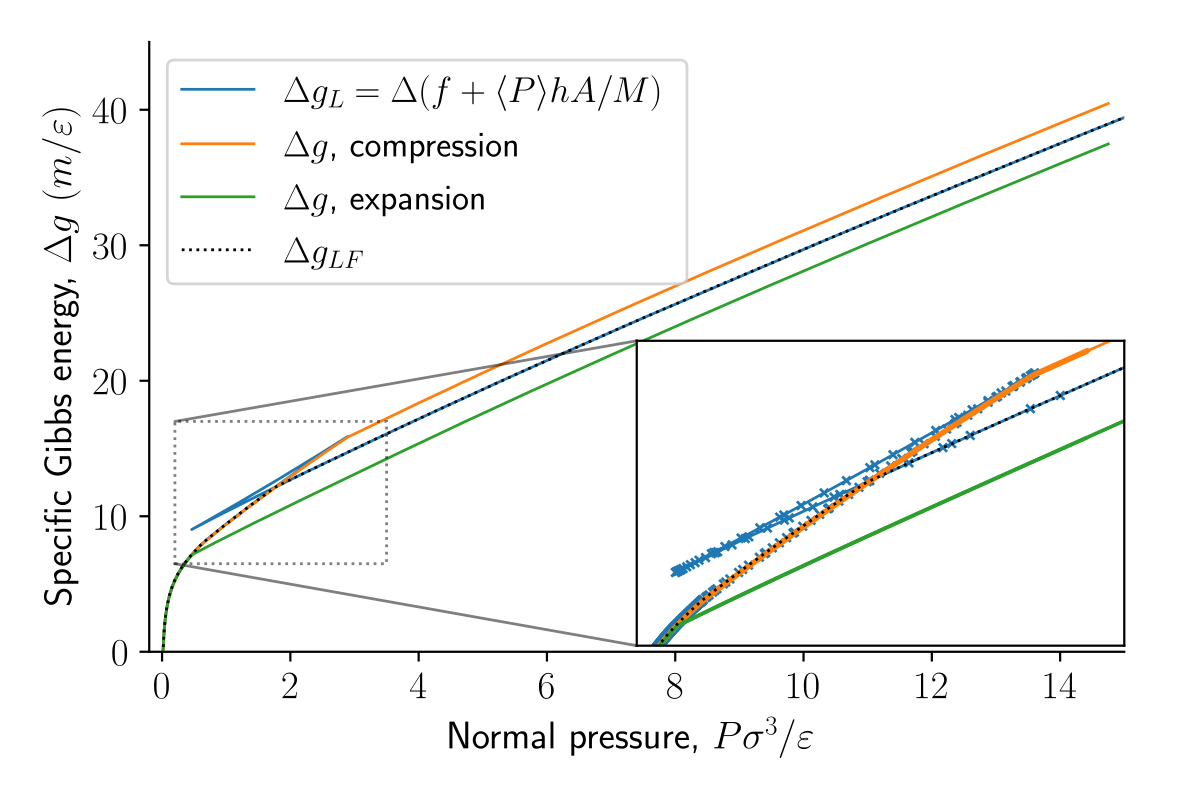}
    \includegraphics[width=\linewidth]{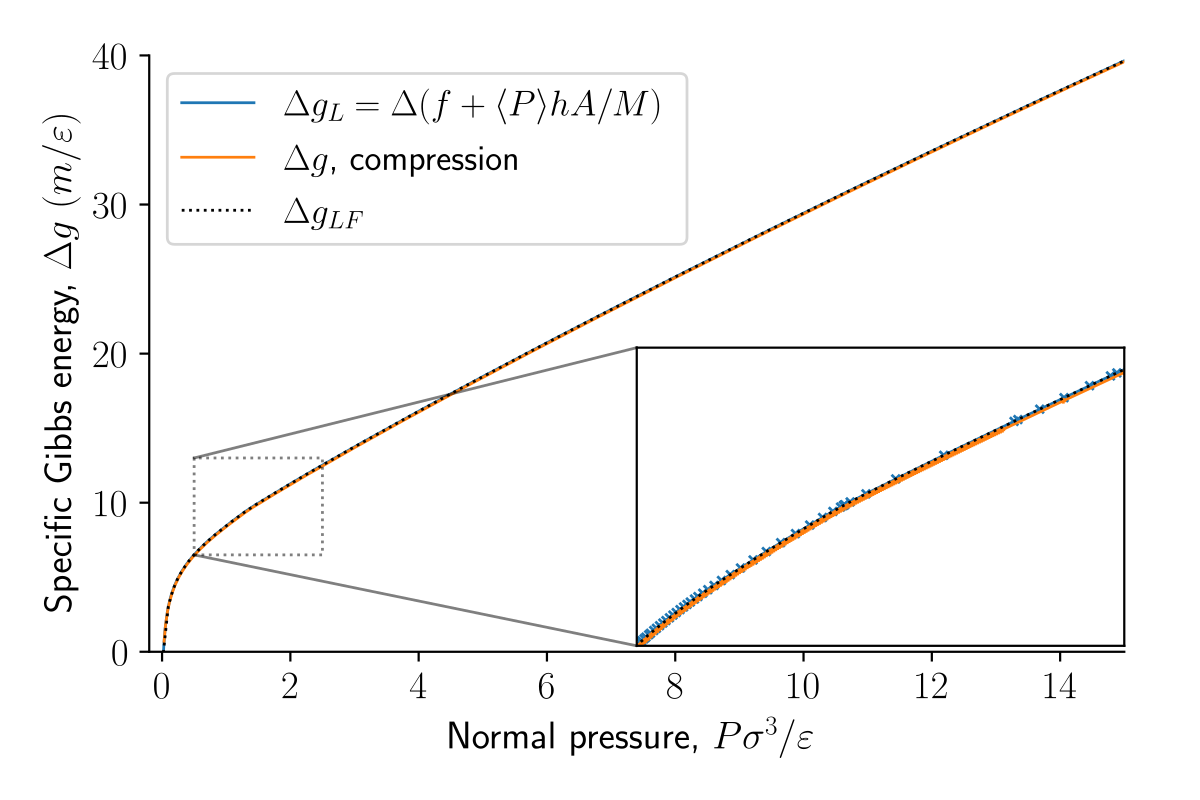}
    \caption{Specific Gibbs energy as a function of normal pressure. The surface number densities from top to bottom are $\Gamma=0.22$, $\Gamma=0.81$, and $\Gamma=1.18$, respectively.}
    \label{fig:free_energy_pressure}
\end{figure}

\begin{figure}
    \centering
    \includegraphics[width=\linewidth]{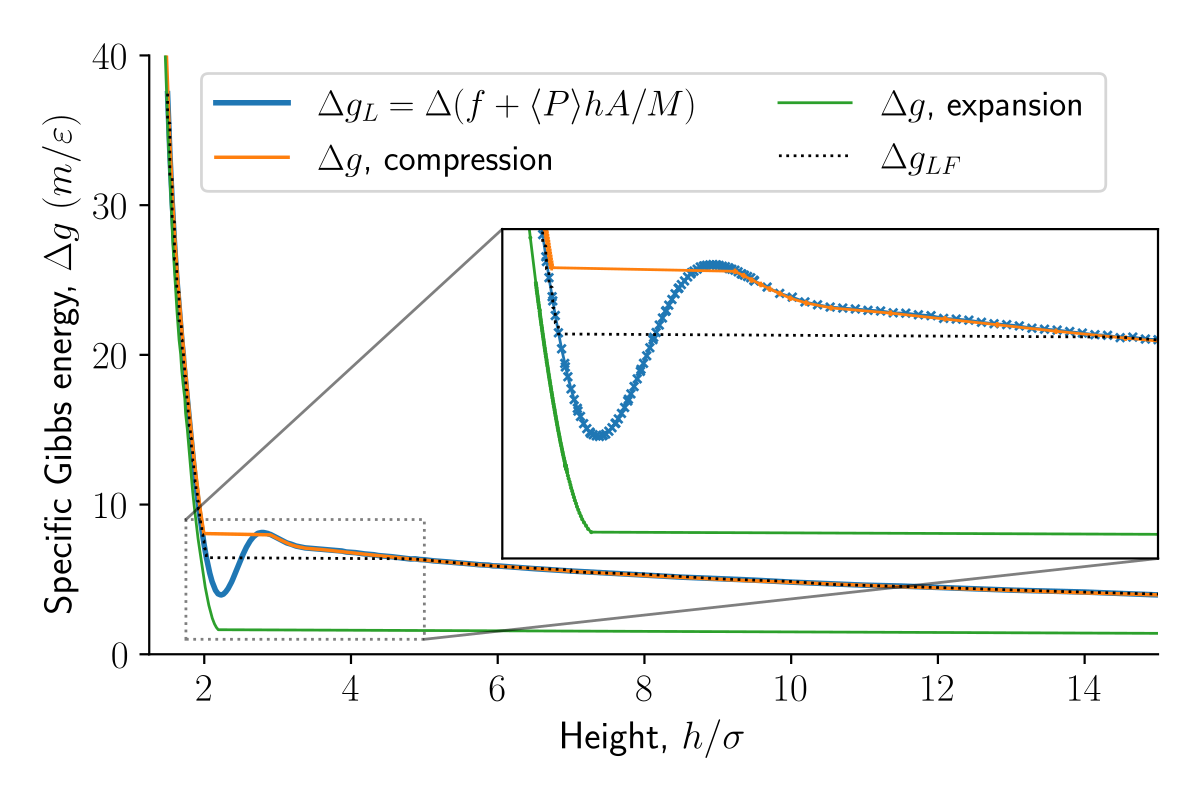}
    \includegraphics[width=\linewidth]{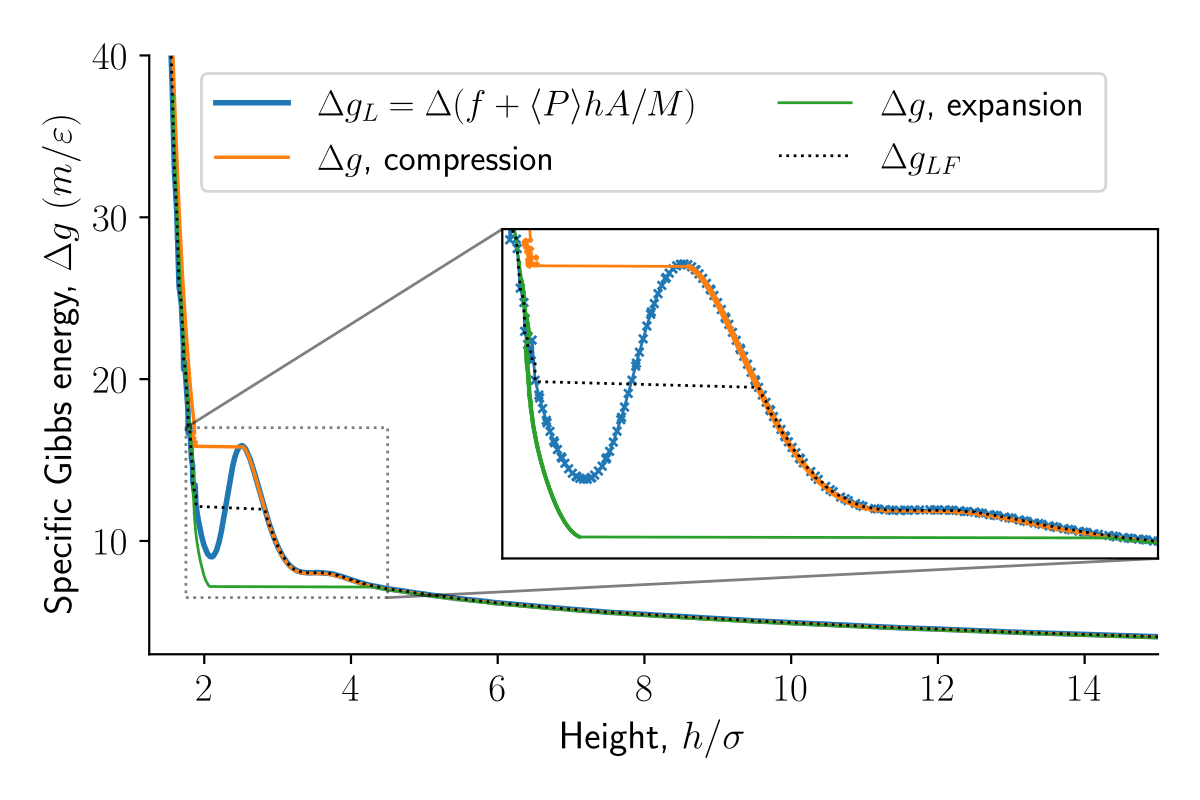}
    \includegraphics[width=\linewidth]{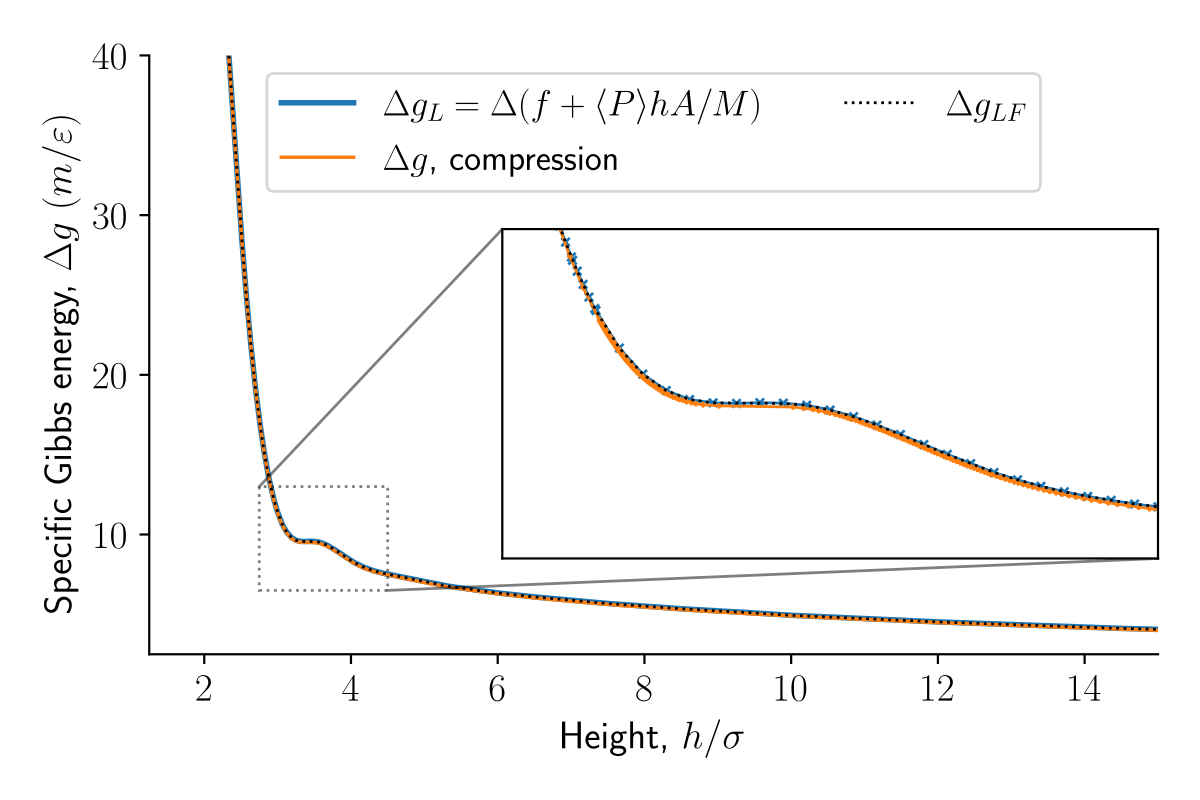}
    \caption{Specific Gibbs energy as a function of height. The surface number density in the figures from top to bottom are $\Gamma=0.22$, $\Gamma=0.81$, and $\Gamma=1.18$, respectively.}
    \label{fig:free_energy_height}
\end{figure}

\subsection{Response functions.}

In the isobaric conditions, it is useful to consider the response function
\begin{equation}
    K_{N,T}=-A\left(\frac{\partial h}{\partial P}\right)_{N,T} = -\left(\frac{\partial^2 G}{\partial P^2}\right)_{N,T} \geq 0,
\end{equation}
where $K_{N,T}/Ah$ is the isothermal compressibility \cite{campa2018concavity}. As the states with negative $K_{N,T}$ cannot be stable under height fluctuations, $K_{N,T}$ is restricted to be non-negative. This can be seen in Fig. \ref{fig:pressure_height}, where the states with negative $K_{N,T}$ in isochoric conditions are not available in isobaric conditions. The region with negative compressibility (positive slope) corresponds to the spinodal region, which is the region between the local minimum and maximum. In the isochoric ensemble, the height is a control variable, and states with negative isothermal compressibility can be realized. In other words, the response function
\begin{equation}
    \frac{1}{K_{N,T}}=-\frac{1}{A}\left(\frac{\partial P}{\partial h}\right)_{N,T} =\frac{1}{A^2} \left(\frac{\partial^2 F}{\partial h^2}\right)_{N,T}
\end{equation}
can be negative. This can be seen in Fig. \ref{fig:pressure_height}.

The function $g_\text{LF}(N,P,T)$ presents two non-smooth points, indicating a first-order phase transition. The isothermal compressibility in isochoric conditions is negative, and is well defined in all available states. In isobaric conditions, the isothermal compressibility is undefined when the free energy is non-smooth. It is well defined and non-negative for all other states. The reason for this is that the normal pressure is increased by applying a directed force that seeks to compress the system. If the system was to increase its volume in response to this compression force, the fluid center of mass would have to move in the direction opposite to the applied force, thus violating conservation of linear momentum. 

\subsection{The specific Gibbs energy as a function of normal pressure. Legendre-Fenchel transforms.}

Figs. \ref{fig:pressure_height} presents the normal pressure-height relationships for isochoric and isobaric conditions. The states available for the system to follow in isochoric conditions are blue, while isobaric conditions provide states given by the orange and green curves. The orange curve is for isobaric compression and the green curve is for isobaric expansion. Once the maximum in a curve is reached during isobaric compression, the system will switch to a smaller height. Alternatively, by isobaric expansion all points near the minimum become accessible. The dotted line is the height computed from the derivative of the Legendre-Fenchel transform $g_\text{LF}$ with respect to the normal pressure $P$, see equation \ref{eq:pressure_def}. This curve gives the Maxwell construction of equal areas, or in other words the straight line across the van der Waals loop or the binodal region. For isobaric conditions, the system is metastable in the binodal region and unstable in the spinodal region. 

The specific Gibbs energy is presented in Figs. \ref{fig:free_energy_pressure} and \ref{fig:free_energy_height} as a function of normal pressure and height, respectively. For the large thermodynamic system, i.e. the bottom panels where $\Gamma = 1.18$, the specific Gibbs energy of compression (orange curve) coincides with the Legendre transform (blue curve), and the Legendre-Fenchel transform (black dotted curve) of the specific Helmholtz energy. This is the expected behavior of large systems.

At lower surface number densities, the Legendre transform ceases to apply. But we can still understand the system in terms of its thermodynamic properties. The normal pressure-height curves form a van der Waals loop for isochoric conditions. The specific Gibbs energy from compression differs from that of expansion for isobaric conditions. The Legendre-Fenchel transform (dotted curve) follows the Legendre transform (blue curve), except in the van der Waals loop, which is cut out by the Legendre-Fenchel transform.

The underlying distribution of normal pressures in the isochoric conditions is highly peaked, implying that the conditions for the saddlepoint approximation, necessary for the Legendre-Fenchel transform, are obeyed. The small error bar in the normal pressure, which is the standard deviation, testifies to this, see Fig. \ref{fig:pressure_height}. The mean relative standard deviation is $1.1$\%, $0.4$\%, $0.09$\%  for $\Gamma=0.22$, $\Gamma=0.81$, and $\Gamma=1.18$, respectively.

The results of the slit pore simulations show that structural changes are more restricted when in isochoric conditions than in isobaric conditions. The findings are similar to observations of polymer stretching \cite{bering2020entropy}. Also here a transition was found between states. But then the different regimes referred to the type of degrees of freedom of the molecule (active or frozen rotational or stretching degrees of freedom) \cite{bering2020entropy, bering2020legendre}. We have seen that the Legendre-Fenchel transforms apply to two widely different cases, so we may pose the question: Does the transform apply in general to energy conversion in small systems? From the mathematics point of view, this seems likely \cite{commons2021duality, qian2021thermodynamic}. More data is needed before we may conclude, but this study brings out an interesting perspective. If the answer is yes, we may have a new tool for energy conversion in small systems. 

When the Helmholtz energy is non-convex in the isochoric ensemble the system exhibits hysteresis in the isobaric ensemble. Hysteresis means that the state of the system depends on the system's history. The specific Gibbs energy depends on whether the system comes from a compressed or expanded state. In the isobaric ensemble, available states can be explored by first compressing and then expanding the system. One of the two available states at the same controlled normal pressure in the compression and expansion curves are metastable, the other state is stable. The Legendre-Fenchel transform gives the stable states. Since the piston cannot deform the energy barrier to go from a metastable state to a stable state is high, which is the reason for the pronounced hysteresis in this system. The loop created in the specific Gibbs energy profile in isobaric conditions is interesting. The existence of a loop means in principle, that work can be extracted from the loop by only two steps, namely compression, and expansion. The slit pore may serve as a very first model for deformable porous media in this context. 

\section{Conclusions}

The normal pressure varies with the height. The excess normal pressure, the normal pressure minus the bulk pressure, is often called the disjoining or solvation pressure \cite{israelachvili2015intermolecular, galteland2021nanothermodynamic}. The disjoining pressure is typically defined in the grand canonical ensemble (an open system). The disjoining pressure oscillates with a period equal to the fluid particle diameter because of the fluid packing between the walls. The mechanism of the excess normal pressure is the same here as for an open system, but the behaviour of the normal pressure observed in the closed system is different than for an open system. The observation of disjoining pressure is not new, see for example the work by Israelachvili \cite{israelachvili2015intermolecular}. It has been observed in simulations as well as experimentally. However, Israelachvili does not apply thermodynamics and the machinery that follows it. In this work, we have expanded upon the work by Israelachvili and others by giving a thermodynamic description. As a result, we get additional relations that can be used, for example, Maxwell relations.

In isochoric conditions, there is a smooth transition from a fluid to a close-packed structure. The Helmholtz energy of this smooth transition is non-convex for small surface number densities. The system is restricted in such a way that there must be a smooth transition. It is not possible for the system to have coexistence of a fluid and a close-packed structure because that would imply that the system could have two different heights at the same time. A possible future generalization could be to allow the solid walls to deform in a less restricted manner such that it would be possible for the system to have varying heights. In this way the system would allow for coexistence of a fluid and a close-packed structure. We hypothesize that the hysteresis in such a system will be less pronounced. Vapor-liquid coexistence is not restricted in this way, the vapor and liquid can have two different volumes. Since the height is controlled, the system is also restricted to that height. However, the normal pressure is controlled in isobaric conditions. As a result, there is a first-order phase transition from a fluid to a close-packed structure for small surface number densities. There is no coexistence of fluid and close-packed structure for the same reason as for isochoric conditions. The Legendre transform does not apply to non-convex free energies, and we have shown the Legendre-Fenchel transform must be applied.

Small systems in Hill's sense are not extensive. They are characterized by giving different responses to their ensemble of control. In the present case, we have studied and compared two small systems, i.e. isochoric and isobaric fluids confined to slit pores. Despite their smallness, we have found that they are perfectly well describable by thermodynamics when the theory is adjusted to deal with smallness.  One such adjustment means to use Legendre-Fenchel transforms, rather than Legendre transforms. Doing that, we have shown that the specific Helmholtz energy can be transformed into the specific Gibbs energy. The findings are general, and support the systematic approach of Hill for descriptions of other small systems. 

\section*{Author Contributions}
O.G. contributed to formal analysis, investigation, methodology, software, and visualisation. D.B. and S.K contributed to supervision. All authors contributed to conceptualization, writing original drafts, reviewing, and editing. All authors have read and agreed to the published version of the manuscript.

\section*{Conflicts of interest}
There are no conflicts to declare.

\section*{Acknowledgements}
The simulations were performed on resources provided by UNINETT Sigma2 - the National Infrastructure for High Performance Computing and Data Storage in Norway. We thank the Research Council of Norway through its Centres of Excellence funding scheme, project number 262644, PoreLab.

\balance
\bibliography{rsc}

\begin{thebibliography}{10}

\bibitem{brennan2002phase}
JK~Brennan and W~Dong.
\newblock {Phase transitions of one-component fluids adsorbed in random porous
  media: Monte Carlo simulations}.
\newblock {\em The Journal of chemical physics}, 116(20):8948--8958, 2002.

\bibitem{brennan2003molecular}
John~K Brennan and Wei Dong.
\newblock {Molecular simulation of the vapor-liquid phase behavior of
  Lennard-Jones mixtures in porous solids}.
\newblock {\em Physical review E}, 67(3):031503, 2003.

\bibitem{galteland2021nanothermodynamic}
Olav Galteland, Dick Bedeaux, and Signe Kjelstrup.
\newblock {Nanothermodynamic description and molecular simulation of a
  single-phase fluid in a slit pore}.
\newblock {\em Nanomaterials}, 11(1):165, 2021.

\bibitem{strom2020thermodynamic}
Bj{\o}rn~A Str{\o}m, Jianying He, Dick Bedeaux, and Signe Kjelstrup.
\newblock {When thermodynamic properties of adsorbed films depend on size:
  Fundamental theory and case study}.
\newblock {\em Nanomaterials}, 10(9):1691, 2020.

\bibitem{strom2021adsorption}
Bj{\o}rn~A Str{\o}m, Dick Bedeaux, and Sondre~K Schnell.
\newblock {Adsorption of an Ideal Gas on a Small Spherical Adsorbent}.
\newblock {\em Nanomaterials}, 11(2):431, 2021.

\bibitem{wilhelmsen2015tolman}
{\O}ivind Wilhelmsen, Dick Bedeaux, and David Reguera.
\newblock {Tolman length and rigidity constants of the Lennard-Jones fluid}.
\newblock {\em The Journal of chemical physics}, 142(6):064706, 2015.

\bibitem{Hill1963}
Terrell~L Hill.
\newblock {\em {Thermodynamics of small systems, part 1}}.
\newblock Benjamin, 1963.

\bibitem{Hill1964}
Terrell~L Hill.
\newblock {\em {Thermodynamics of small systems, part 2}}.
\newblock Benjamin, 1964.

\bibitem{dong2021thermodynamics}
W.~Dong.
\newblock {Thermodynamics of interfaces extended to nanoscales by introducing
  integral and differential surface tensions}.
\newblock {\em Proceedings of the National Academy of Sciences}, 118(3), 2021.

\bibitem{kjelstrup2018non}
Signe Kjelstrup, Dick Bedeaux, Alex Hansen, Bj{\o}rn Hafskjold, and Olav
  Galteland.
\newblock {Non-isothermal transport of multi-phase fluids in porous media. the
  entropy production}.
\newblock {\em Frontiers in Physics}, 6:126, 2018.

\bibitem{kjelstrup2019non}
Signe Kjelstrup, Dick Bedeaux, Alex Hansen, Bj{\o}rn Hafskjold, and Olav
  Galteland.
\newblock {Non-isothermal transport of multi-phase fluids in porous media.
  Constitutive equations}.
\newblock {\em Frontiers in Physics}, 6:150, 2019.

\bibitem{Bedeaux2020}
Dick Bedeaux, Signe Kjelstrup, and Sondre~K Schnell.
\newblock {\em {Nanothermodynamics. General theory}}.
\newblock NTNU, Trondheim, Norway, 2020.

\bibitem{galteland2019pressures}
Olav Galteland, Dick Bedeaux, Bj{\o}rn Hafskjold, and Signe Kjelstrup.
\newblock {Pressures inside a nano-porous medium. The case of a single phase
  fluid}.
\newblock {\em Frontiers in Physics}, 7:60, 2019.

\bibitem{erdHos2020gibbs}
M{\'a}t{\'e} Erd{\H{o}}s, Olav Galteland, Dick Bedeaux, Signe Kjelstrup,
  Othonas~A Moultos, and Thijs~JH Vlugt.
\newblock {Gibbs ensemble Monte Carlo simulation of fluids in confinement:
  Relation between the differential and integral pressures}.
\newblock {\em Nanomaterials}, 10(2):293, 2020.

\bibitem{rauter2020two}
Michael~T Rauter, Olav Galteland, M{\'a}t{\'e} Erd{\H{o}}s, Othonas~A Moultos,
  Thijs~JH Vlugt, Sondre~K Schnell, Dick Bedeaux, and Signe Kjelstrup.
\newblock {Two-phase equilibrium conditions in nanopores}.
\newblock {\em Nanomaterials}, 10(4):608, 2020.

\bibitem{bering2020entropy}
Eivind Bering, Signe Kjelstrup, Dick Bedeaux, J~Miguel Rubi, and Astrid~S
  de~Wijn.
\newblock {Entropy production beyond the thermodynamic limit from
  single-molecule stretching simulations}.
\newblock {\em The Journal of Physical Chemistry B}, 124(40):8909--8917, 2020.

\bibitem{bering2020legendre}
Eivind Bering, Dick Bedeaux, Signe Kjelstrup, Astrid~S de~Wijn, Ivan Latella,
  and J~Miguel Rubi.
\newblock {A Legendre--Fenchel Transform for Molecular Stretching Energies}.
\newblock {\em Nanomaterials}, 10(12):2355, 2020.

\bibitem{galteland2020solvent}
Olav Galteland, Fernando Bresme, and Bj{\o}rn Hafskjold.
\newblock {Solvent-Mediated Forces between Ellipsoidal Nanoparticles Adsorbed
  at Liquid--Vapor Interfaces}.
\newblock {\em Langmuir}, 36(48):14530--14538, 2020.

\bibitem{israelachvili2015intermolecular}
Jacob~N Israelachvili.
\newblock {\em {Intermolecular and surface forces}}.
\newblock Academic press, 2015.

\bibitem{touchette2005legendre}
Hugo Touchette.
\newblock {Legendre-Fenchel transforms in a nutshell}, 2005.
\newblock available at: \url{
  https://appliedmaths.sun.ac.za/\~htouchette/archive/notes/lfth2.pdf} (Nov.
  2021).

\bibitem{rockafellar2015convex}
Ralph~Tyrell Rockafellar.
\newblock {\em {Convex analysis}}.
\newblock Princeton university press, 2015.

\bibitem{plimpton1995fast}
Steve Plimpton.
\newblock {Fast parallel algorithms for short-range molecular dynamics}.
\newblock {\em Journal of computational physics}, 117(1):1--19, 1995.

\bibitem{thompson2021lammps}
Aidan~P Thompson, H~Metin Aktulga, Richard Berger, Dan~S Bolintineanu,
  W~Michael Brown, Paul~S Crozier, Pieter~J in't Veld, Axel Kohlmeyer, Stan~G
  Moore, Trung~Dac Nguyen, et~al.
\newblock {LAMMPS-A flexible simulation tool for particle-based materials
  modeling at the atomic, meso, and continuum scales}.
\newblock {\em Computer Physics Communications}, page 108171, 2021.

\bibitem{hafskjold2019thermodynamic}
Bj{\o}rn Hafskjold, Karl~Patrick Travis, Amanda~Bailey Hass, Morten Hammer,
  Ailo Aasen, and {\O}ivind Wilhelmsen.
\newblock {Thermodynamic properties of the 3D Lennard-Jones/spline model}.
\newblock {\em Molecular Physics}, 117(23-24):3754--3769, 2019.

\bibitem{stukowski2009visualization}
Alexander Stukowski.
\newblock {Visualization and analysis of atomistic simulation data with
  OVITO--the Open Visualization Tool}.
\newblock {\em Modelling and Simulation in Materials Science and Engineering},
  18(1):015012, 2009.

\bibitem{kristiansen2020transport}
Kim~R. Kristiansen.
\newblock {Transport Properties of the Simple Lennard-Jones/Spline Fluid I:
  Binary Scattering and High-Accuracy Low-Density Transport Coefficients}.
\newblock {\em Frontiers in Physics}, 8:271, 2020.

\bibitem{campa2018concavity}
Alessandro Campa, Lapo Casetti, Ivan Latella, Agust{\'\i}n P{\'e}rez-Madrid,
  and Stefano Ruffo.
\newblock {Concavity, response functions and replica energy}.
\newblock {\em Entropy}, 20(12):907, 2018.

\bibitem{commons2021duality}
Jeffrey Commons, Ying-Jen Yang, and Hong Qian.
\newblock {Duality Symmetry, Two Entropy Functions, and an Eigenvalue Problem
  in Gibbs' Theory}.
\newblock {\em arXiv preprint arXiv:2108.08948}, 2021.

\bibitem{qian2021thermodynamic}
Hong Qian.
\newblock {Thermodynamic Behavior of Statistical Event Counting in Time:
  Independent and Correlated Measurements}.
\newblock {\em arXiv preprint arXiv:2109.12806}, 2021.

\end{thebibliography}
\bibliographystyle{unsrt}
\end{document}